\documentclass[aip,jmp,amsmath,amssymb,preprint,]{revtex4-1}

\usepackage{graphicx}% Include figure files
\usepackage{dcolumn}% Align table columns on decimal point
\usepackage{bm}% bold math
\usepackage{cancel}
\usepackage{amsmath}
\usepackage{extarrows}
\usepackage[version=3]{mhchem}
\usepackage{booktabs}

\begin{document}

%\preprint{AIP/123-QED}

\title[Chemical reaction networks under spatial confinement and crowding conditions]
{Brownian dynamics simulations of an idealized chemical reaction network under spatial confinement and crowding conditions}
%\thanks{Footnote to title of article.}

\author{Benjamin B. Bales}
\affiliation{ 
Department of Computer Science,
University of California Santa Barbara \\
93106 Santa Barbara, CA}
\author{Giovanni Bellesia}
\affiliation{ 
Los Alamos National Laboratory,
87545 Los Alamos, NM}
\email{giovanni.bellesia@gmail.com}

\date{\today}

\begin{abstract}
234

%Valid PACS numbers may be entered using the \verb+\pacs{#1}+ command.
\end{abstract}
% PACS, the Physics and Astronomy % Classification Scheme.

\maketitle

\section{\label{sec:intro}Introduction}
\begin{figure}[!ht]
\centering
\includegraphics[scale=0.5]{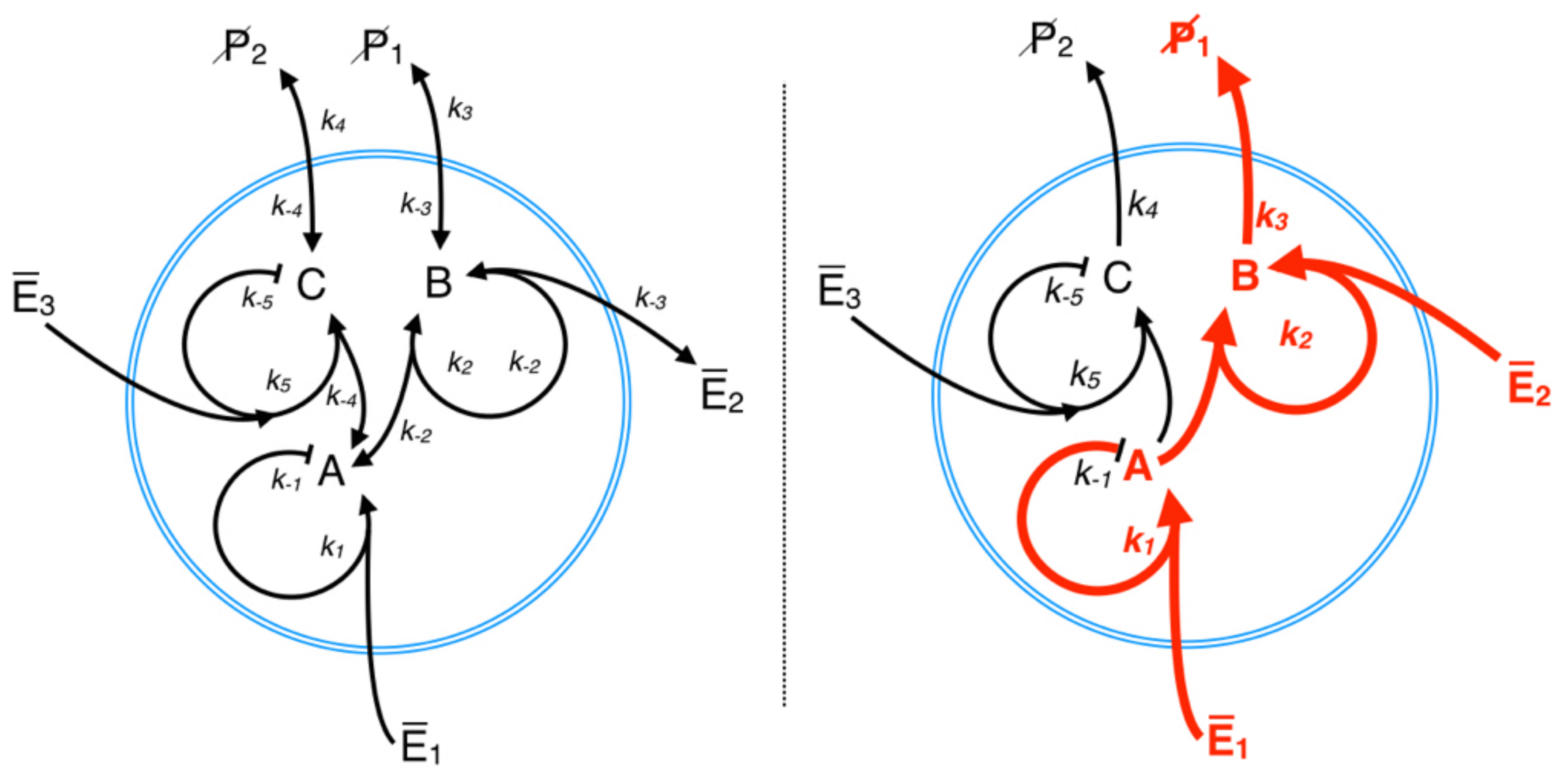}
\caption{Schematic view of the Willamowski-Rossler chemical network. (Left) the \textit{full} version. (Right) the \textit{minimal} version analyzed in this study, obtained by setting $k_{-2}=k_{-3}=k_{-4}=0$. In red we highlight the Lotka-Volterra component of the network. The rate equations for both versions exhibit three different dynamical regimes: fixed point, limit cycle and chaotic  \cite{Willamowski1980, Stucki2005}.}
\label{fig:1}
\end{figure}

Biochemical networks \textit{in vivo} are typically \textit{open} to the exchange of energy and matter with the surrounding environment\cite{Qian2006,Qian2007,Ritort2008,Stano2013}. They often contain autocatalytic steps \cite{IUPAC,Ghadiri1997,Dadon2008,Ikegami2013,Steel2010} and their dynamics tends to be strongly influenced by thermal and intrinsic noise \cite{Johnson1987,Gillespie1977}, macromolecular crowding and spatial confinement \cite{Minton1981,Minton1998,Minton2001,Lichter2007,Lichter2008}. In this study we present a simple computational model of a generic biochemical network \textit{in vivo} and we investigate how its dynamics is affected by spatial confinement and particle crowding. \cite{Minton1981,Minton1998,Minton2001,Lichter2007,Lichter2008}.  Our model is based on the Willamowski-Rossler (WR) chemical network \cite{Willamowski1980}. The WR network (see Figure \ref{fig:1}(a)) is a non-linear chemical system based on zeroth, first and second order chemical reactions. It contains three autocatalytic steps involving species $A$,$B$ and $C$ and it is thermodynamically open \cite{Qian2006,Qian2007,Ritort2008,Stano2013}. Its rate equations display a rich and complicated dynamics comprising fixed point, limit cycle and chaotic attractors. The WR network has been previously studied via deterministic and non-spatial stochastic simulation methods \cite{Clarke1988,Matias1993,Nicolis1993,Kapral1996,Kapral2002,Stucki2005} but never as a stochastic reaction--diffusion system where crowding and spatial confinement are explicitly taken into account. 

In detail, we investigate the effects of spatial confinement and crowding on a \textit{minimal} version of the WR network (MWR) (see Figure \ref{fig:1}(b) and Ref.~\onlinecite{Stucki2005}) using hard-sphere \cite{Ando2010,Skolnick2011} Brownian dynamics simulations integrating chemical reactivity \cite{Morelli2008,Frazier2012}. We fix the population numbers for species $E_1$, $E_2$, $E_3$, $P_1$ and $P_2$ (consequently the rates $k_1$, $k_3$ and $k_5$ become pseudo-first order) so that the MWR network is thermodynamically open. The following chemical reactions describe the MWR system used in our simulations \cite{Stucki2005} (see also Figure \ref{fig:1}(b)).
\begin{subequations}
\label{eq:1}
\begin{align}
\cee{\bar{E}_1 + A &<=>[\text{$k_1$}][\text{$k_{-1}$}] 2A} \\
\cee{A + B & ->[\text{$k_2$}] 2B} \\
\cee{A + C & ->[\text{$k_4$}] \cancel{P}_2} \\
\cee{\bar{E}_2 + B & ->[\text{$k_3$}] \cancel{P}_1} \\
\cee{\bar{E}_3 + C & <=>[\text{$k_5$}][\text{$k_{-5}$}] 2C}.
\end{align}
\end{subequations}

The main assumption in the MWR system \cite{Willamowski1980,Stucki2005} is that three of the backward reaction rate constants shown in Figure \ref{fig:1}(a), namely $k_{-2}$, $k_{-3}$ and $k_{-4}$, are much smaller than their forward counterparts and, hence, can be neglected (See Figure \ref{fig:1}(b)). The MWR system is composed by two main subsystems: a Lotka-Volterra oscillator \cite{Lotka1910,Lotka1920,Volterra1926} involving species $A$ and $B$ and a chemical \textit{switch} \cite{Clarke1988} that couples the Lotka-Volterra component to species $C$. Similarly to the `full' WR network, the MWR rate equations derived from the set of chemical reactions $(1a)-(1e)$ display a diverse dynamical behavior comprising fixed point, limit cycle and chaotic attractors \cite{Willamowski1980,Stucki2005}. 
We quantify the effects of confinement and crowding on the population dynamics, flux of information and spatial organization within the MWR network. Our approach and analysis can be naturally extended to more complicated chemical networks and can be potentially relevant to a number of open problems in biochemistry such as the synthesis of primitive cellular units (protocells) and the definition of their role in the chemical origin of life, the characterization of vesicle-mediated drug delivery processes and, more generally,  the study of biochemical networks \textit{in vivo} \cite{Rasmussen2008,Stano2013,Luisi2006,Szath2005}. We make the case for a more widespread development and use of spatial stochastic simulation methods for biochemical networks \textit{in vivo} that explicitly take into account confinement and macromolecular crowding \cite{Beck2011,Linda2013,Andersen2005,Minton1981}. 

\section{\label{sec:meth}Methods}
All three autocatalytic species $A$, $B$ and $C$ are spatially confined within a spherical \textit{container}, $E_1$ and $E_3$ catalyze the synthesis of $A$, and $C$, respectively, whereas $E_2$ catalyzes the degradation of $C$. $P_1$ and $P_2$ are the products of reactions $(1d)$ and $(1c)$, respectively, and they get instantaneously eliminated from the reaction pool, i.e., their constant population number is zero. The constant population numbers of $E_1$, $E_2$ and $E_3$ and the instantaneous elimination of $P_1$ and $P_2$ lead to a biochemical network composed by $A$, $B$ and $C$ which is spatially enclosed and thermodynamically open, i.e., it exchanges matter and energy with the surrounding environment by means of three \textit{sources} ($E_1$, $E_2$ and $E_3$) and two \textit{sinks} ($P_1$ and $P_2$). The constant values of $E_1$, $E_2$, $E_3$ are incorporated into the \textit{pseudo}-first order rates $k_1$, $k_3$, $k_5$, respectively (see Figure \ref{fig:1}).

The different chemical species in the MWR system are modeled as \textit{reactive}, Brownian hard spheres confined in a spherical container. The details of the Brownian integrator used in our simulations can be found in Refs.~\onlinecite{Morelli2008,Frazier2012}. The radius of the hard spheres for species $A$, $B$, and $C$ is $0.01$ $\mu m$ and the diffusion coefficient is $D=0.01$ $\mu m^2 s^{-1}$. In all our simulations the time step is fixed at $\Delta t = 0.01$ $s$. 

To study the effects of crowding and confinement we run two separate sets of reactive Brownian dynamics simulations. In the first set we consider six different spherical containers with radius varying between $0.4$ and 0.65  $\mu m$. The containers are implemented as `hard-wall' spherical boundary conditions. For each of the six spherical containers we run a total of $30$ independent simulations, each of total time $t_{tot}=1000$ $s$. Three sets of values for the reaction rate constants (\textit{kset1}, \textit{kset2}, \textit{kset3}) are used for each one of the six different spherical containers. They correspond to three distinct dynamical behaviors in the deterministic implementation of the MWR model: fixed point, limit cycle and chaotic dynamics, respectively. The first set (\textit{kset1}, fixed point attractor) is $k_1=30.0$, $k_{-1}= 0.25$, $k_2 = 1.0$, $k_3=10.0$, $k_4 = 0.4$, $k_5=16.5$,  $k_{-5} = 0.5$. To generate the second set (\textit{kset2} - limit cycle attractor) we simply consider the first set of parameters and change the value of $k_4$ to $0.6$. In the third set (\textit{kset3} - chaotic attractor) we set $k_4=0.6$, $k_5=18.5$ and $k^{-5}=0.4$. In other words, \textit{kset2} is generated from \textit{kset1} by increasing the degradation of A and C (increasing the coupling between the Lotka-Volterra component and the \textit{switch}) while \textit{kset3} is obtained from \textit{kset1} by increasing both the $A-C$ coupling and decreasing the ratio $k_{-5}/k_{5}$. 
We run $10$ independent simulations for each of the three parameter sets. 
%The units for the reaction rates are $\mu m^{-3} s^{-1}$, $s^{-1}$ and $\mu m^{3}s^{-1}$ for zeroth, first and second order reactions respectively. 
The starting point for each simulation is generated randomly placing $A=B=C=100$ hard spheres within the proper spherical container. In the second set of simulations we take into account the presence of a variable number of `chemically inert' crowders modeled as hard spheres of radius $r=0.01$ $\mu m$ and with diffusion coefficient $D=0.01$ $\mu m^2 s^{-1}$. The starting point for each simulation in the second set is generated randomly placing $A=B=C=100$ hard spheres  and a variable number of inert crowders in a spherical container with radius  $R=0.4$ $\mu m$. We run independent simulations for five different crowder population numbers: varying between $2\times10^3$ and $8\times10^3$. For each of the five crowder population numbers we run a total of $30$ independent simulations ($10$ for each of the three parameters sets), each of total time $t_{tot}=1000$ $s$. 

\section{\label{sec:res}Results and Discussion}
\subsection{Population dynamics}
\begin{figure}[!ht]
\centering
\includegraphics[scale=0.5]{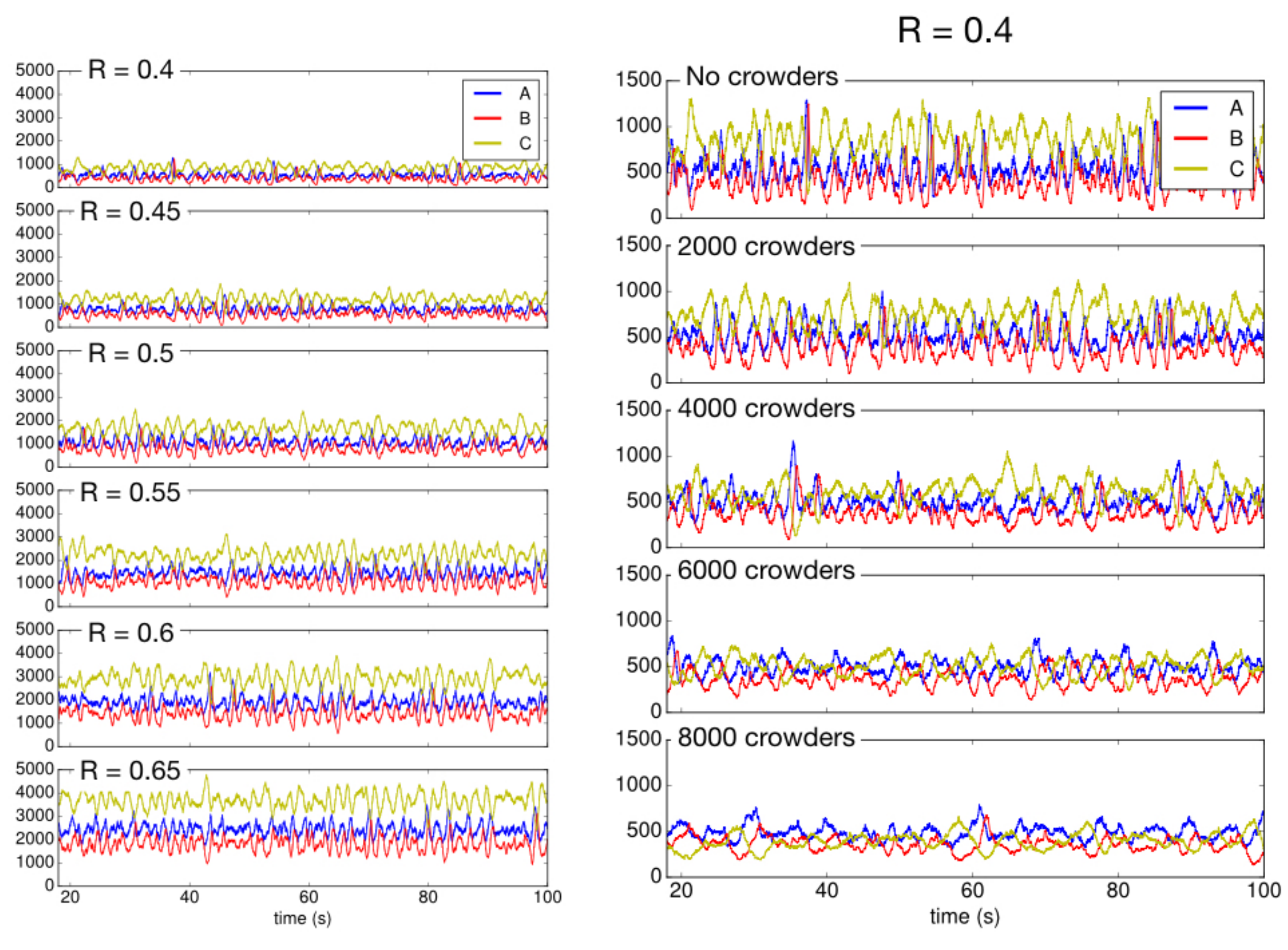}
\caption{Population time series (partial time window) obtained each from a single trajectory of our brownian dynamics simulations of the MWR network with \textit{kset3} parameterization. Left panel: population time series for systems with varying volume (radii varying from $0.4$ to $0.65$ $\mu$m).
Right panel: population time series for systems with varying number of crowders and with constant container volume (radius is $0.4$ $\mu$m).}
\label{fig:2}
\end{figure}

\begin{figure}[!ht]
\centering
\includegraphics[scale=0.6]{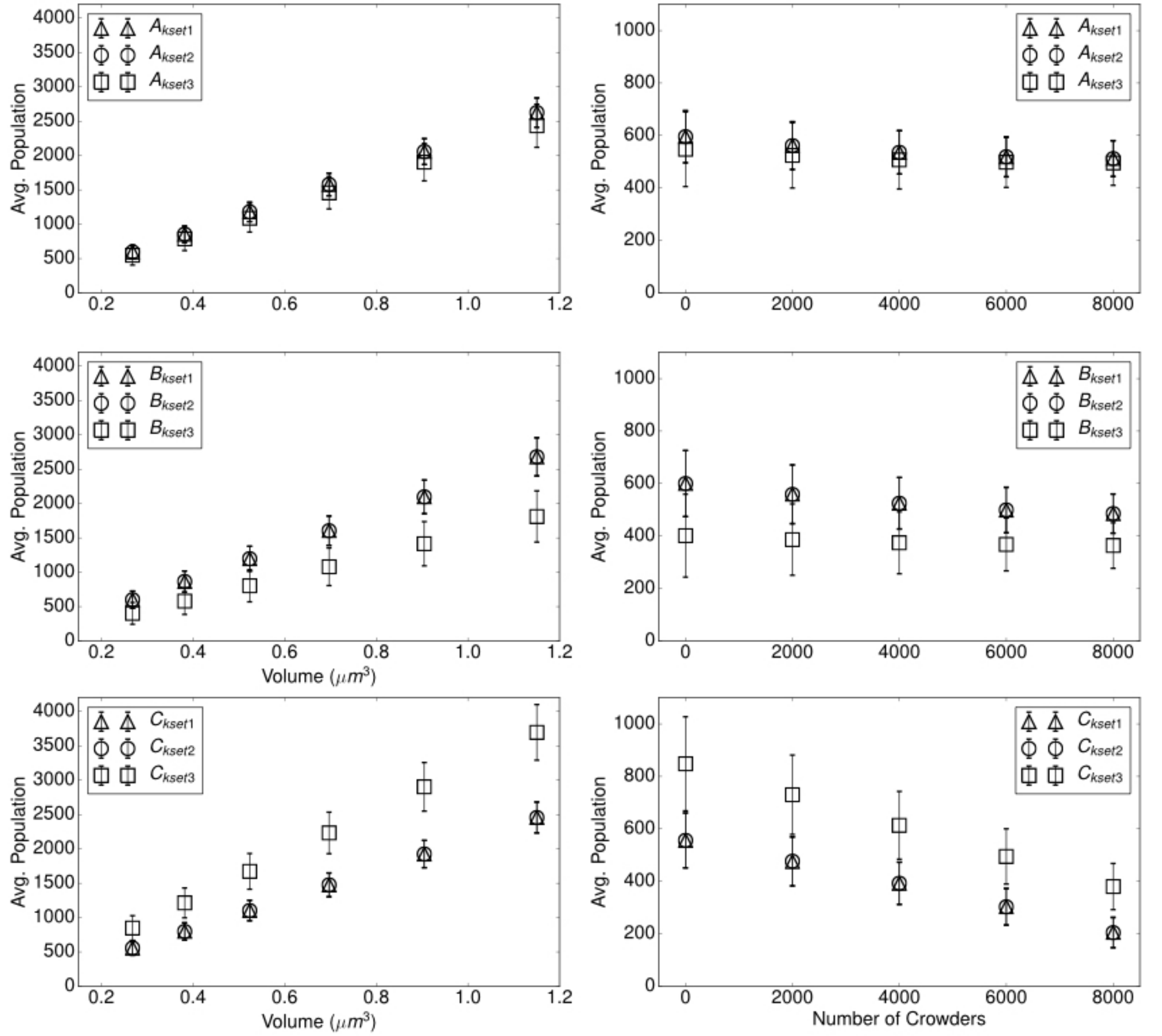}
\caption{Average population values for species $A$ (top), $B$ (middle) and $C$ (bottom) with parameterizations \textit{kset1}, \textit{kset2} and \textit{kset3}.
Left panel: the average populations are plotted against the volume of the spherical container, no inert crowders are present. The average population of the reactive species grows linearly with the volume of the spherical container . Right panel: the average populations are plotted against the number of inert crowders. A slight decrease in the average population is observed for larger crowders numbers.
}
\label{fig:3}
\end{figure}

We focus our analysis on the stationary \cite{VanKampen2007} portion of our Brownian simulations.
In Figure \ref{fig:2} we show a set of representative time windows for the population numbers of species $A$, $B$ and $C$ related to simulations with variable container volume (left panel), and to simulations with constant container volume and varying number of inert crowders (right panel). All data refer to parameterization \textit{kset3} (see Section \ref{sec:meth}). Time series population data generated under \textit{kset1} and \textit{kset2} are not shown as they display analogous temporal patterns.
Figure \ref{fig:2} qualitatively shows that (1) both average population and fluctuations increase with increasing container volume for all species and (2) the presence of an increasing number of inert crowders affects the average population of species $A$, $B$ and $C$. It also appears to lower both the fluctuations in the population dynamics and the temporal interdependence between the different species.
A quantitative assessment of the mean and fluctuations dependence from both the container volume and the crowders number is given in Figure \ref{fig:3}. In the left panel we show that in the limited range of container volumes considered in our simulations, the mean population increases linearly with increasing container volume. The fluctuations calculated as the standard deviation from the mean also have a tendency to increase although the actual functional dependency is not immediately clear. 
The effects of the presence of inert crowders are shown in the right panel of Figure \ref{fig:3}. All species show a decrease in their average population for increasing crowders numbers which can be intuitively related to the diminished availability of free volume within the spherical container.
An additional observation on the data in Figure \ref{fig:3} relates to the dependence of the average population from the parameterization set. First, the population dynamics of species $A$, $B$ and $C$ does not change significantly when the parameterization set changes from \textit{kset1} to \textit{kset2}. Second, the transition from parameterization \textit{kset1} and \textit{kset2} to \textit{kset3} has opposite effects on species $B$ and $C$. Third, species $A$ does not show any quantifiable dependence from the parameter set (under both volume and crowders' number varying conditions). It is easy to connect the increase in the slope of the $C$(volume) linear fit to the increase in $C$'s net synthesis going from  \textit{kset1} and \textit{kset2} to \textit{kset3}. The decrease in the linear fit's slope for species $B$ is less clear since species $B$ is not directly affected by the changes in the parameterization set and species $A$, which is directly coupled to $B$, is insensitive to those changes. The insensitivity to parameter changes in species $A$ can be qualitatively explained considering that $A$ is the connection point in the MWR network between the Lotka-Volterra component and the \textit{switch} component \cite{Clarke1988} and therefore benefits from the `modulation' given by the interaction with both species $B$ and $C$.

\subsection{Information flux and statistical complexity}
\begin{figure}[!ht]
\centering
\includegraphics[scale=0.6]{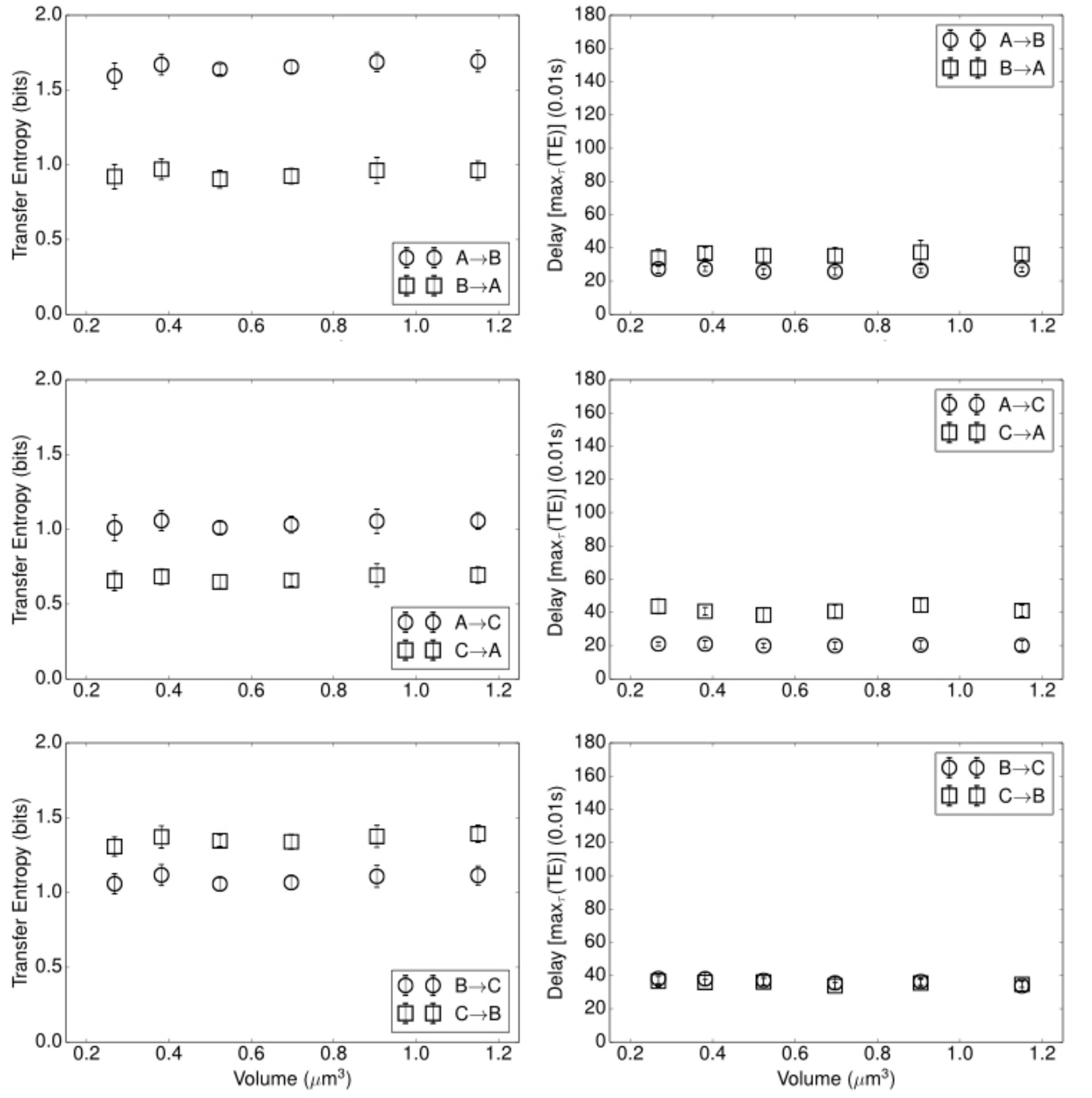}
\caption{Left panel: transfer entropy as a function of the container volume. Right panel: characteristic time delay correspondent to the maximum value of the transfer entropy. Data refer to \textit{kset3} parameterization.}
\label{fig:4}
\end{figure}

\begin{figure}[!ht]
\centering
\includegraphics[scale=0.6]{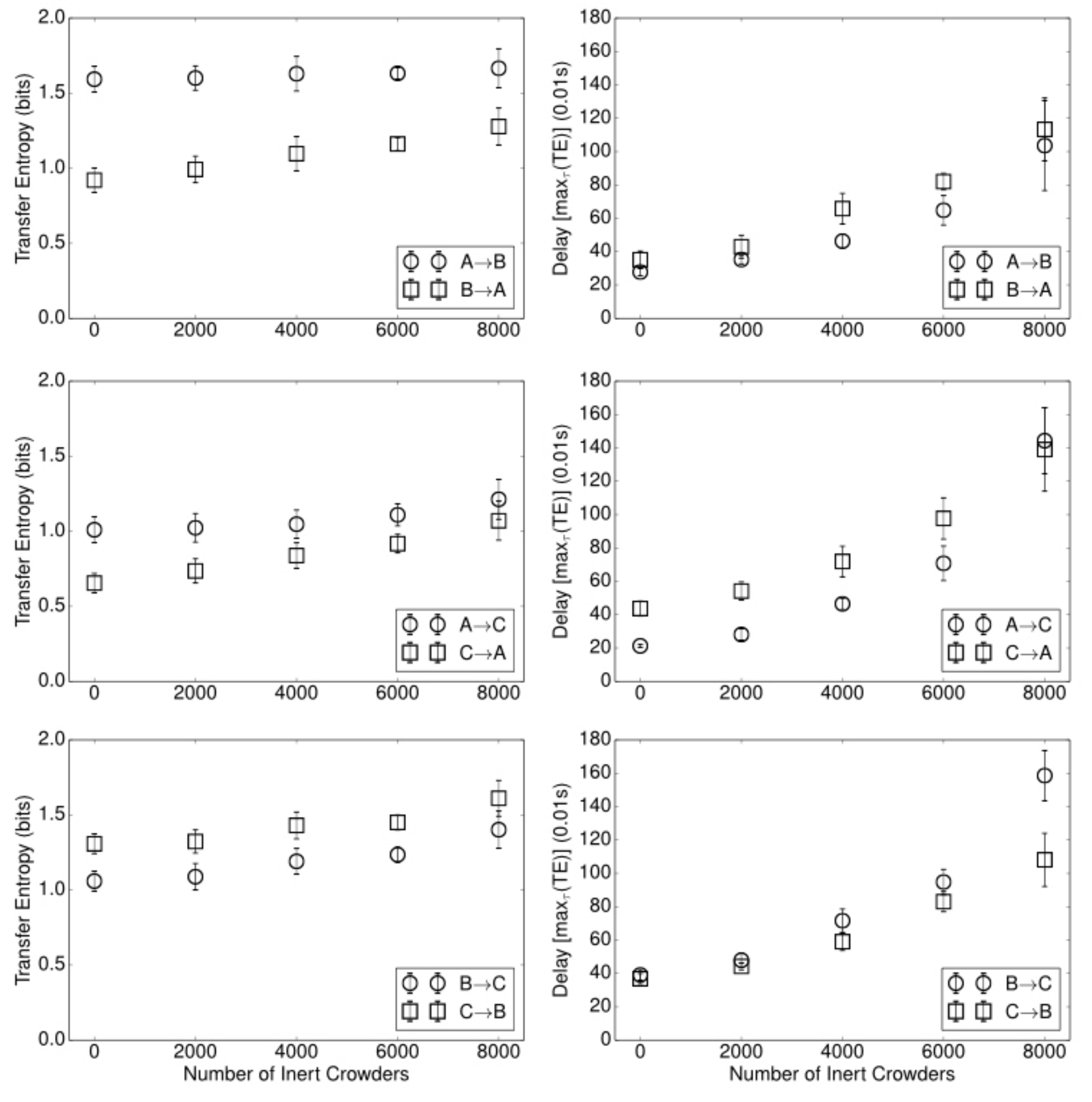}
\caption{Left panel: transfer entropy as a function of the number of inert crowders. Right panel: characteristic time delay correspondent to the maximum value of the transfer entropy. Data refer to \textit{kset3} parameterization.}
\label{fig:5}
\end{figure}

A possible explanation of the peculiar behavior of species $A$ and its relation with $A$'s `double coupling' within the MWR network comes from the analysis of the information flux quantified by the transfer entropy defined as a particular case of the conditional mutual information \cite{Schreiber2000,Palus2007,Palus2008}:
\begin{equation}
\label{eq:2}
\max_{\tau} I(Y(t),X(t+\tau) | X(t)),
\end{equation}
where $I(X,Y | Z)$ is the mutual information between $X$ and $Y$ given $Z$ \cite{Cover2006} and $\tau$ is the time delay. We employ transfer entropy to estimate both the amount and the direction of the information flux in the MWR network. The value of the delay parameter $\tau$ considered in our calculations corresponds to the maximum of the transfer entropy for a given trajectory pair. 
A number of interesting conclusions can be inferred from the analysis of the transfer entropy data in Figures \ref{fig:4} and \ref{fig:5}. Considering first the chemical network as a whole it is worth noting that the varying container volume does not significantly affect the information flux between the different species in the network. 
For systems with variable number of inert crowders there is a small but noticeable systematic increase in the transfer entropy with differences between the less and the most crowded systems of the order of $0.3-0.4$ bits. A further look at the behavior of the single species shows the pivotal role of species $A$ as a common \textit{influencer} of the dynamics of species $B$ and $C$. Both Figure \ref{fig:4} and \ref{fig:5} show that the amount of information transferred from species $A$ is systematically larger than the information transferred to species $A$ in both volume-varying and crowding number-varying systems. This asymmetry in the information flux (common to all three parameterization sets \textit{kset1}, \textit{kset2} and \textit{kset3} - data not shown) can be linked to an increased ability of $A$ to `absorb' external perturbations and therefore to its lower sensitivity to parameter change (see previous Section).  The main difference in terms of information transfer between systems with and without inert crowders (see Figures \ref{fig:4} and \ref{fig:5}, respectively) is in the characteristic time delay $\tau$ (see Equation \ref{eq:2}) at which the information transfer is maximal. While the characteristic delay $\tau$ is not affected by changes in the container volume (right panel in Figure \ref{fig:4}), the presence of an increasing number of inert crowders in a constant volume container decreases the speed at which the information is transferred within the network (right panel in Figure \ref{fig:5}).

In order to improve the clarity and conciseness of our manuscript, from now on we focus only on simulations performed under parameterization set \textit{kset3} as this set of parameters seems to have an additional layer of complexity with respect to \textit{kset1} and \textit{kset2} (data not shown) and carries all the significant information about our system. 
Information theory functionals can be also used to estimate the degree of complexity in the time evolution of the chemical network and its dependence from the container volume and from the presence of crowders. The complexity estimation quantity that we choose is an intensive statistical complexity measure which is the product of the normalized spectral entropy $\hat{S}(P_r)$ and the intensive Jensen-Shannon divergence 
$\hat{Q}(P_r,P_e)$ \cite{Percival1979,Fuentes2007} defined respectively as:

\begin{equation}
\label{eq:3}
\hat{S(P_r)} = -S_0\sum_{r'}^{N_f} P_{r'} \log_2 P_{r'}
\end{equation}

with

\begin{equation}
P_r = \frac{f_r^2}{ \sum_{r'}^{N_f} f_{r'}^2},
\end{equation}
where $f_r$ are the frequencies in the Fourier spectrum and $N_f=4000$ is the number of frequencies considered, and

\begin{equation}
\label{eq:4}
\hat{Q}(P_r,P_e) = Q_0 \Big[ S\Big(\frac{P_r + P_e}{2}\Big) - \frac{1}{2}S(P_e) - \frac{1}{2}S(P_r)\Big],
\end{equation}

where $S(P_e) = \log_2 N_f=S_0^{-1}$ and $Q_0$ is the normalization factor for $Q$ and $P_e=1/N_f$.

\begin{figure}[!ht]
\centering
\includegraphics[scale=0.6,angle=-90]{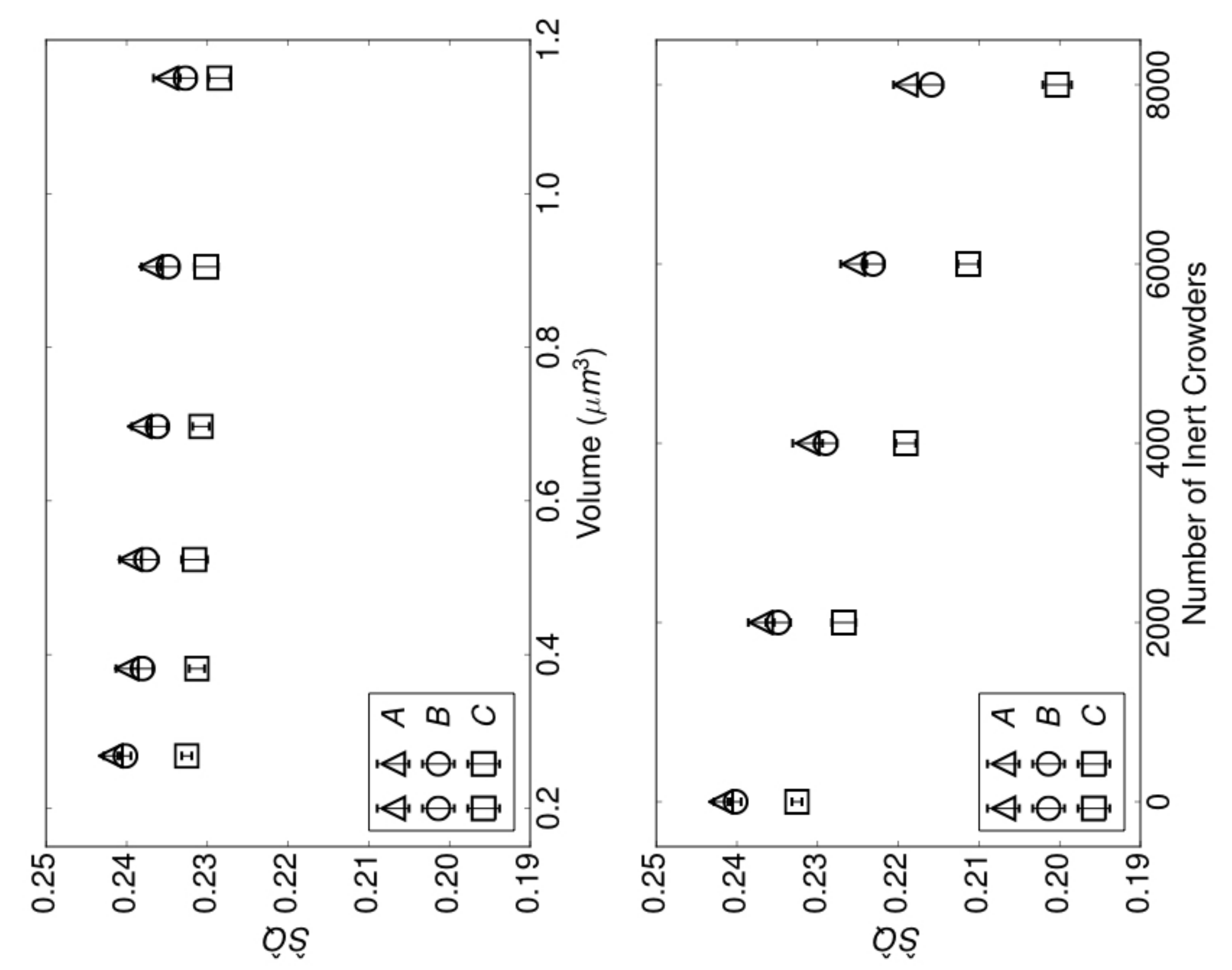}
\caption{Statistical complexity measure as a function of the container volume (top) and number of inert crowders (bottom). Data refer to species $A$, $B$, $C$ under \textit{kset3} parameterization.}
\label{fig:6}
\end{figure}

The statistical complexity $\hat{S}\hat{Q}$ is zero for both $P_r = \{1,0,0,\cdots,0\}$ and $P_r = P_e = 1/N_f$, i.e., for spectral entropy $S=0$ and $S=\log_2 N_f$ (fully ordered and fully stochastic systems) \cite{Fuentes2007}. The results for the statistical complexity $\hat{S}\hat{Q}$ are shown in Figure \ref{fig:6}. The top panel shows that container volume variability does not significantly affect the average statistical complexity for species $A$, $B$ and $C$ (both $\hat{S} and \hat{Q}$ do not vary significantly - $\\Delta leq 0.02$). Conversely, for systems with constant volume and variable crowders number the statistical complexity decreases with increasing number of crowders. In detail, the decrease is almost exclusively due to a decrease in the normalized spectral entropy from $0.62$ to $0.55$, $0.61$ to $0.54$ and $0.58$ to $0.50$, for species $A$, $B$ and $C$, respectively. The intensive Jensen-Shannon divergence remains constant at around $0.39-0.40$. As a general conclusion from our information theoretic analysis, we can state that the presence of a growing number of inert crowders drives the chemical network toward a lower degree of complexity which is possibly driven by a more efficient information transfer Figure \ref{fig:5}) between the reactive chemical species.

\subsection{Spatial statistics}
The spatial organization of the chemical species in the network and its coupling with their population sizes are investigated employing a deterministic implementation of the DBSCAN clustering algorithm \cite{Ester1996}, where `boundary' particles are discarded as noise and with parameterization 
$\epsilon = 0.06 \mu$m and $n_c=4$. $\epsilon$ is the cutoff distance defining particle pairs belonging to the same cluster and therefore `connected' to each other, and $n_c$ is the minimum number of `connections' that defines a `core' particle \cite{Ester1996}. The presence of inert crowders strongly influences the spatial organization of the chemical species $A$, $B$ and $C$ in the MWR network. 

\begin{figure}[!ht]
\centering
\includegraphics[scale=0.6,angle=-90]{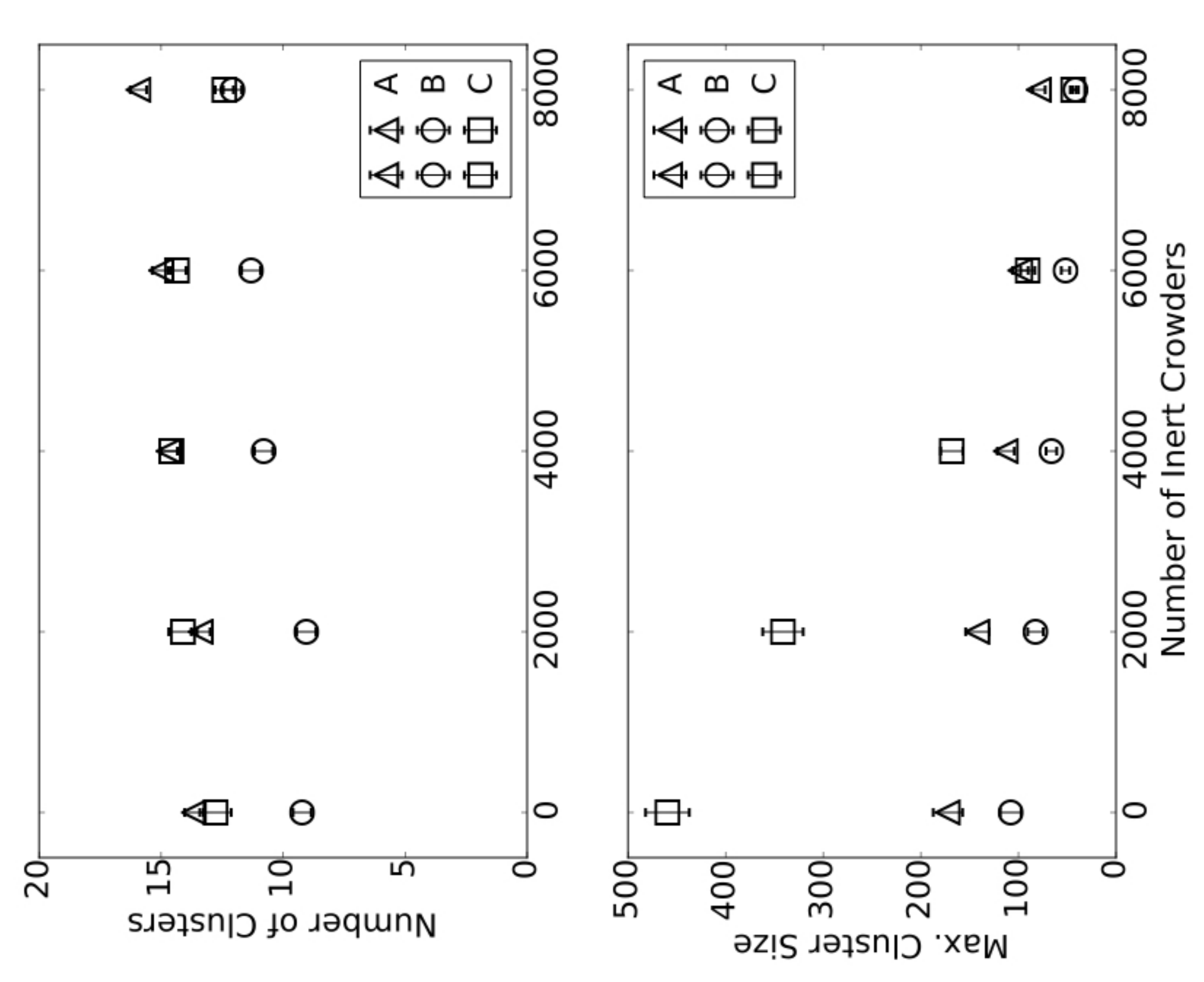}
\caption{Average number of clusters (top) and average maximum cluster size (bottom) as a function of the number of inert clusters. Data refer to species $A$, $B$, $C$ under \textit{kset3} parameterization. The average number of clusters shows a weak tendency to increase for all three chemical species. The average maximum cluster size decreases with denser crowding conditions. The max. cluster size in species $C$ displays the largest decrease rate.}
\label{fig:7}
\end{figure}

\begin{figure}[!ht]
\centering
\includegraphics[scale=0.65]{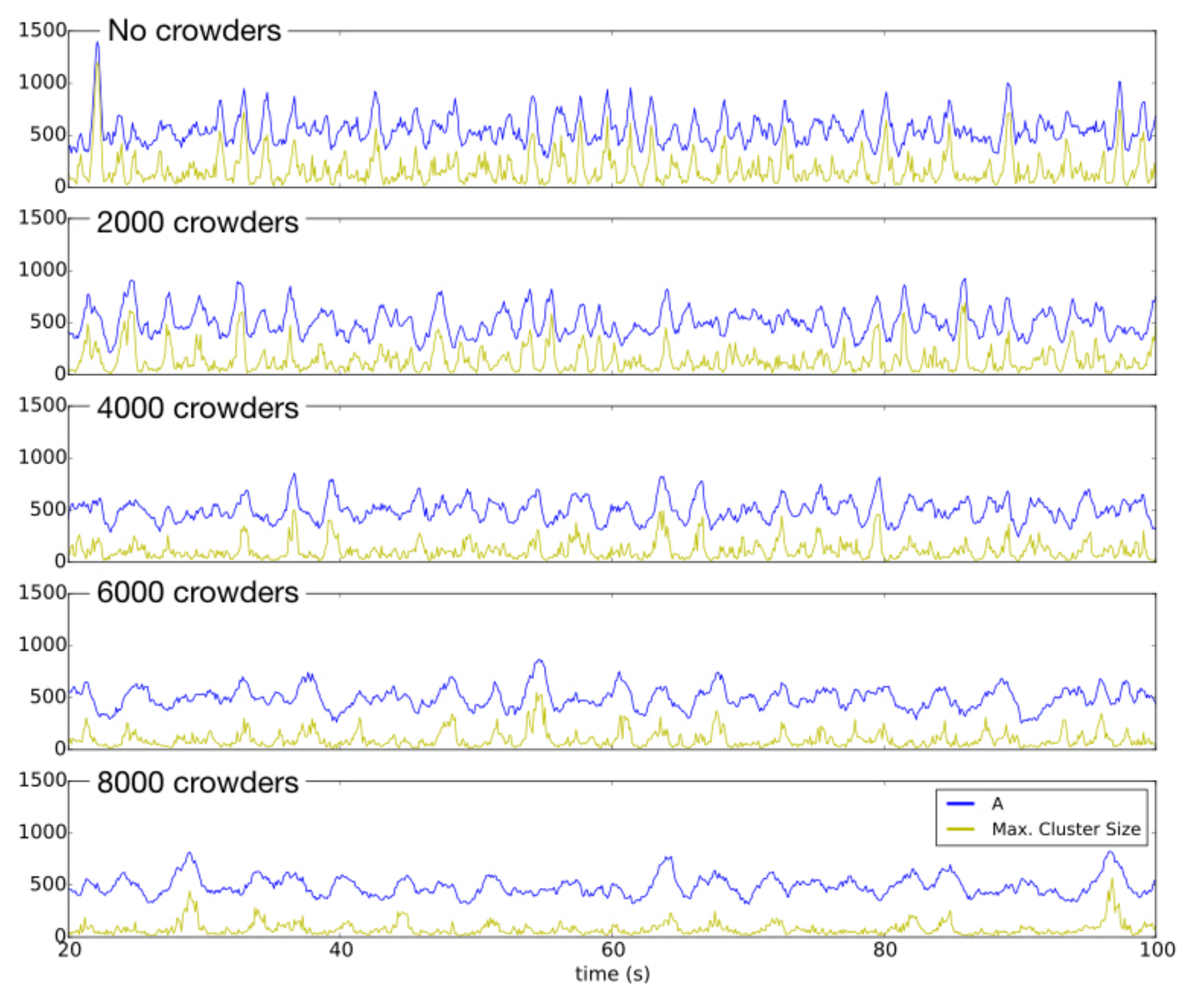}
\caption{Time evolution for the population size (blue) and the maximum cluster size (yellow). The temporal pattern in the maximum cluster size accurately mirrors the population size. Data refer to species $A$ under \textit{kset3} parameterization. The mutual information between the two sets of temporal data decreases with increasing number of crowders (see Table \ref{tab:1}).}
\label{fig:8}
\end{figure}

In Figure \ref{fig:7} we show the average number of clusters (top) and average maximum cluster size (bottom) as a function of the number of inert clusters. On the one hand, the average number of clusters shows a weak tendency to increase for all three chemical species. On the other hand, the average maximum cluster size decreases with denser crowding conditions. Among the three reactive species the maximum cluster size in species $C$ displays both the largest values and the largest decrease rate. Figure \ref{fig:7} basically shows that the presence of an increasing number of crowders opposes the natural tendency of the reactive particles in our system to accumulate in well-defined regions of the available space. An interesting feature of the maximum cluster size temporal evolution is shown in Figure \ref{fig:8}. For small numbers of crowders the maximum cluster size for species $A$ tightly mirrors the time evolution of the population of species $A$ (species $B$ and $C$ show very similar behavior - data not shown). The `correlation' between population dynamics and maximum cluster dynamics weakens with increasing crowders number. Indeed, table \ref{tab:1} shows that the mutual information \cite{Cover2006} between population and maximum cluster dynamics decreases with increasing crowders numbers. 

\section{\label{sec:conc}Conclusions}
In this study we investigate the dynamical behavior of a simple chemical network under spatial confinement and crowding. We observe that the presence of inert crowders affects in a non-trivial way the population dynamics of the reactive species in the network. The detailed analysis of the population dynamics of the MWR network under different confinement and crowding conditions presented in Section \ref{sec:res} represents, from a more general perspective, an example of the level of detail, not accessible to deterministic and stochastic well-mixed models, that can be resolved when spatial confinement and crowding are explicitly taken into account. In conclusion, we try to make the case for the use of spatial stochastic simulations as an elective method to complement experiments and to improve our understanding of complex systems where dynamics is both spatially confined and compartmentalized.

\begin{acknowledgments}
The authors would like to thank Linda Petzold, Zachary Frazier, Frank Alber, Marco J. Morelli and Steve Plimpton for useful discussions and for the help with the testing of our Brownian simulator. This work has been supported by NIH grant 1R01EB014877-01.
\end{acknowledgments}

\bibliographystyle{apsrev}
\bibliography{crowd1}

\begin{thebibliography}{44}
\expandafter\ifx\csname natexlab\endcsname\relax\def\natexlab#1{#1}\fi
\expandafter\ifx\csname bibnamefont\endcsname\relax
  \def\bibnamefont#1{#1}\fi
\expandafter\ifx\csname bibfnamefont\endcsname\relax
  \def\bibfnamefont#1{#1}\fi
\expandafter\ifx\csname citenamefont\endcsname\relax
  \def\citenamefont#1{#1}\fi
\expandafter\ifx\csname url\endcsname\relax
  \def\url#1{\texttt{#1}}\fi
\expandafter\ifx\csname urlprefix\endcsname\relax\def\urlprefix{URL }\fi
\providecommand{\bibinfo}[2]{#2}
\providecommand{\eprint}[2][]{\url{#2}}

\bibitem[{\citenamefont{Qian}(2006)}]{Qian2006}
\bibinfo{author}{\bibfnamefont{H.}~\bibnamefont{Qian}}, \bibinfo{journal}{J
  Phys Chem B} \textbf{\bibinfo{volume}{110}}, \bibinfo{pages}{15063}
  (\bibinfo{year}{2006}).

\bibitem[{\citenamefont{Qian}(2007)}]{Qian2007}
\bibinfo{author}{\bibfnamefont{H.}~\bibnamefont{Qian}}, \bibinfo{journal}{Annu
  Rev Phys Chem} \textbf{\bibinfo{volume}{58}}, \bibinfo{pages}{113}
  (\bibinfo{year}{2007}).

\bibitem[{\citenamefont{Ritort}(2008)}]{Ritort2008}
\bibinfo{author}{\bibfnamefont{F.}~\bibnamefont{Ritort}},
  \emph{\bibinfo{title}{Nonequilibrium fluctuations in small systems: from
  physics to biology}} (\bibinfo{publisher}{John Wiley and Sons, Inc.},
  \bibinfo{year}{2008}), vol. \bibinfo{volume}{137},
  chap.~\bibinfo{chapter}{2}, pp. \bibinfo{pages}{31--123}.

\bibitem[{\citenamefont{Stano and Luisi}(2013)}]{Stano2013}
\bibinfo{author}{\bibfnamefont{P.}~\bibnamefont{Stano}} \bibnamefont{and}
  \bibinfo{author}{\bibfnamefont{P.~L.} \bibnamefont{Luisi}},
  \bibinfo{journal}{Curr Opin Biotechnol} \textbf{\bibinfo{volume}{24}},
  \bibinfo{pages}{633} (\bibinfo{year}{2013}).

\bibitem[{\citenamefont{McNaught and Wilkinson}(1997)}]{IUPAC}
\bibinfo{author}{\bibfnamefont{A.~D.} \bibnamefont{McNaught}} \bibnamefont{and}
  \bibinfo{author}{\bibfnamefont{A.}~\bibnamefont{Wilkinson}},
  \emph{\bibinfo{title}{IUPAC. Compendium of Chemical Terminology}}
  (\bibinfo{publisher}{Blackwell Scientific Publications, Oxford},
  \bibinfo{year}{1997}), \bibinfo{edition}{2nd} ed.

\bibitem[{\citenamefont{Lee et~al.}(1997)\citenamefont{Lee, Severin, and
  Ghadiri}}]{Ghadiri1997}
\bibinfo{author}{\bibfnamefont{D.~H.} \bibnamefont{Lee}},
  \bibinfo{author}{\bibfnamefont{K.}~\bibnamefont{Severin}}, \bibnamefont{and}
  \bibinfo{author}{\bibfnamefont{M.~R.} \bibnamefont{Ghadiri}},
  \bibinfo{journal}{Current Opinion in Chemical Biology}
  \textbf{\bibinfo{volume}{1}}, \bibinfo{pages}{491} (\bibinfo{year}{1997}).

\bibitem[{\citenamefont{Dadon et~al.}(2008)\citenamefont{Dadon, Wagner, and
  Ashkenasy}}]{Dadon2008}
\bibinfo{author}{\bibfnamefont{Z.}~\bibnamefont{Dadon}},
  \bibinfo{author}{\bibfnamefont{N.}~\bibnamefont{Wagner}}, \bibnamefont{and}
  \bibinfo{author}{\bibfnamefont{G.}~\bibnamefont{Ashkenasy}},
  \bibinfo{journal}{Angew Chem Int Ed Engl} \textbf{\bibinfo{volume}{47}},
  \bibinfo{pages}{6128} (\bibinfo{year}{2008}).

\bibitem[{\citenamefont{Virgo and Ikegami}(2013)}]{Ikegami2013}
\bibinfo{author}{\bibfnamefont{N.}~\bibnamefont{Virgo}} \bibnamefont{and}
  \bibinfo{author}{\bibfnamefont{T.}~\bibnamefont{Ikegami}}, in
  \emph{\bibinfo{booktitle}{ECAL - General Track}} (\bibinfo{year}{2013}).

\bibitem[{\citenamefont{Hordijk et~al.}(2010)\citenamefont{Hordijk, Hein, and
  Steel}}]{Steel2010}
\bibinfo{author}{\bibfnamefont{W.}~\bibnamefont{Hordijk}},
  \bibinfo{author}{\bibfnamefont{J.}~\bibnamefont{Hein}}, \bibnamefont{and}
  \bibinfo{author}{\bibfnamefont{M.}~\bibnamefont{Steel}},
  \bibinfo{journal}{Entropy} \textbf{\bibinfo{volume}{12}},
  \bibinfo{pages}{1733} (\bibinfo{year}{2010}).

\bibitem[{\citenamefont{Johnson}(1987)}]{Johnson1987}
\bibinfo{author}{\bibfnamefont{H.~A.} \bibnamefont{Johnson}},
  \bibinfo{journal}{Q Rev Biol} \textbf{\bibinfo{volume}{62}},
  \bibinfo{pages}{141} (\bibinfo{year}{1987}).

\bibitem[{\citenamefont{Gillespie}(1977)}]{Gillespie1977}
\bibinfo{author}{\bibfnamefont{D.~T.} \bibnamefont{Gillespie}},
  \bibinfo{journal}{The Journal of Physical Chemistry}
  \textbf{\bibinfo{volume}{81}}, \bibinfo{pages}{2340} (\bibinfo{year}{1977}).

\bibitem[{\citenamefont{Minton}(1981)}]{Minton1981}
\bibinfo{author}{\bibfnamefont{A.~P.} \bibnamefont{Minton}},
  \bibinfo{journal}{Biopolymers} \textbf{\bibinfo{volume}{20}},
  \bibinfo{pages}{2093} (\bibinfo{year}{1981}).

\bibitem[{\citenamefont{Minton}(1998)}]{Minton1998}
\bibinfo{author}{\bibfnamefont{A.~P.} \bibnamefont{Minton}},
  \bibinfo{journal}{Methods in Enzymology} \textbf{\bibinfo{volume}{295}},
  \bibinfo{pages}{127} (\bibinfo{year}{1998}).

\bibitem[{\citenamefont{Minton}(2001)}]{Minton2001}
\bibinfo{author}{\bibfnamefont{A.~P.} \bibnamefont{Minton}},
  \bibinfo{journal}{Journal of Biological Chemistry}
  \textbf{\bibinfo{volume}{276}}, \bibinfo{pages}{10577}
  (\bibinfo{year}{2001}).

\bibitem[{\citenamefont{Richter et~al.}(2007)\citenamefont{Richter, Nessling,
  and Lichter}}]{Lichter2007}
\bibinfo{author}{\bibfnamefont{K.}~\bibnamefont{Richter}},
  \bibinfo{author}{\bibfnamefont{M.}~\bibnamefont{Nessling}}, \bibnamefont{and}
  \bibinfo{author}{\bibfnamefont{P.}~\bibnamefont{Lichter}},
  \bibinfo{journal}{J Cell Sci} \textbf{\bibinfo{volume}{120}},
  \bibinfo{pages}{1673} (\bibinfo{year}{2007}).

\bibitem[{\citenamefont{Richter et~al.}(2008)\citenamefont{Richter, Nessling,
  and Lichter}}]{Lichter2008}
\bibinfo{author}{\bibfnamefont{K.}~\bibnamefont{Richter}},
  \bibinfo{author}{\bibfnamefont{M.}~\bibnamefont{Nessling}}, \bibnamefont{and}
  \bibinfo{author}{\bibfnamefont{P.}~\bibnamefont{Lichter}},
  \bibinfo{journal}{Biochim Biophys Acta} \textbf{\bibinfo{volume}{1783}},
  \bibinfo{pages}{2100} (\bibinfo{year}{2008}).

\bibitem[{\citenamefont{Willamowski and Rossler}(1980)}]{Willamowski1980}
\bibinfo{author}{\bibfnamefont{K.~D.} \bibnamefont{Willamowski}}
  \bibnamefont{and} \bibinfo{author}{\bibfnamefont{O.}~\bibnamefont{Rossler}},
  \bibinfo{journal}{Zeitschrift fur Naturforschung}
  \textbf{\bibinfo{volume}{35a}}, \bibinfo{pages}{317} (\bibinfo{year}{1980}).

\bibitem[{\citenamefont{Aguda and Clarke}(1988)}]{Clarke1988}
\bibinfo{author}{\bibfnamefont{B.~D.} \bibnamefont{Aguda}} \bibnamefont{and}
  \bibinfo{author}{\bibfnamefont{B.~L.} \bibnamefont{Clarke}},
  \bibinfo{journal}{The Journal of Chemical Physics}
  \textbf{\bibinfo{volume}{89}}, \bibinfo{pages}{7428} (\bibinfo{year}{1988}).

\bibitem[{\citenamefont{Guemez and Matias}(1993)}]{Matias1993}
\bibinfo{author}{\bibfnamefont{J.}~\bibnamefont{Guemez}} \bibnamefont{and}
  \bibinfo{author}{\bibfnamefont{M.~A.} \bibnamefont{Matias}},
  \bibinfo{journal}{Physical Review E} \textbf{\bibinfo{volume}{48}},
  \bibinfo{pages}{R2351} (\bibinfo{year}{1993}).

\bibitem[{\citenamefont{Geysermans and Nicolis}(1993)}]{Nicolis1993}
\bibinfo{author}{\bibfnamefont{P.}~\bibnamefont{Geysermans}} \bibnamefont{and}
  \bibinfo{author}{\bibfnamefont{G.}~\bibnamefont{Nicolis}},
  \bibinfo{journal}{The Journal of Chemical Physics}
  \textbf{\bibinfo{volume}{99}}, \bibinfo{pages}{8964} (\bibinfo{year}{1993}).

\bibitem[{\citenamefont{Goryachev and Kapral}(1996)}]{Kapral1996}
\bibinfo{author}{\bibfnamefont{A.}~\bibnamefont{Goryachev}} \bibnamefont{and}
  \bibinfo{author}{\bibfnamefont{R.}~\bibnamefont{Kapral}},
  \bibinfo{journal}{Physical Review Letters} \textbf{\bibinfo{volume}{78}},
  \bibinfo{pages}{1619} (\bibinfo{year}{1996}).

\bibitem[{\citenamefont{Chavez and Kapral}(2002)}]{Kapral2002}
\bibinfo{author}{\bibfnamefont{F.}~\bibnamefont{Chavez}} \bibnamefont{and}
  \bibinfo{author}{\bibfnamefont{R.}~\bibnamefont{Kapral}},
  \bibinfo{journal}{Physical Review E} \textbf{\bibinfo{volume}{65}},
  \bibinfo{pages}{056203} (\bibinfo{year}{2002}).

\bibitem[{\citenamefont{Stucki and Urbanczik}(2005)}]{Stucki2005}
\bibinfo{author}{\bibfnamefont{J.~W.} \bibnamefont{Stucki}} \bibnamefont{and}
  \bibinfo{author}{\bibfnamefont{R.}~\bibnamefont{Urbanczik}},
  \bibinfo{journal}{Zeitschrift fur Naturforschung}
  \textbf{\bibinfo{volume}{60a}}, \bibinfo{pages}{599} (\bibinfo{year}{2005}).

\bibitem[{\citenamefont{Ando and Skolnick}(2010)}]{Ando2010}
\bibinfo{author}{\bibfnamefont{T.}~\bibnamefont{Ando}} \bibnamefont{and}
  \bibinfo{author}{\bibfnamefont{J.}~\bibnamefont{Skolnick}},
  \bibinfo{journal}{Proceedings of the National Academy of Sciences}
  \textbf{\bibinfo{volume}{107}}, \bibinfo{pages}{18457}
  (\bibinfo{year}{2010}).

\bibitem[{\citenamefont{Ando and Jeffrey}(2011)}]{Skolnick2011}
\bibinfo{author}{\bibfnamefont{T.}~\bibnamefont{Ando}} \bibnamefont{and}
  \bibinfo{author}{\bibfnamefont{S.}~\bibnamefont{Jeffrey}}, in
  \emph{\bibinfo{booktitle}{Proceedings of the International Conference of the
  Quantum Bio-Informatics IV}} (\bibinfo{publisher}{Georgia Institute of
  Technology, World Scientific Publishing}, \bibinfo{year}{2011}),
  vol.~\bibinfo{volume}{28}, pp. \bibinfo{pages}{413--426}.

\bibitem[{\citenamefont{Morelli and ten Wolde}(2008)}]{Morelli2008}
\bibinfo{author}{\bibfnamefont{M.~J.} \bibnamefont{Morelli}} \bibnamefont{and}
  \bibinfo{author}{\bibfnamefont{P.~R.} \bibnamefont{ten Wolde}},
  \bibinfo{journal}{J Chem Phys} \textbf{\bibinfo{volume}{129}},
  \bibinfo{pages}{054112} (\bibinfo{year}{2008}).

\bibitem[{\citenamefont{Frazier and Alber}(2012)}]{Frazier2012}
\bibinfo{author}{\bibfnamefont{Z.}~\bibnamefont{Frazier}} \bibnamefont{and}
  \bibinfo{author}{\bibfnamefont{F.}~\bibnamefont{Alber}}, \bibinfo{journal}{J
  Comput Biol} \textbf{\bibinfo{volume}{19}}, \bibinfo{pages}{606}
  (\bibinfo{year}{2012}).

\bibitem[{\citenamefont{Lotka}(1910)}]{Lotka1910}
\bibinfo{author}{\bibfnamefont{J.~A.} \bibnamefont{Lotka}},
  \bibinfo{journal}{The Journal of Chemical Physics}
  \textbf{\bibinfo{volume}{14}}, \bibinfo{pages}{271} (\bibinfo{year}{1910}).

\bibitem[{\citenamefont{Lotka}(1920)}]{Lotka1920}
\bibinfo{author}{\bibfnamefont{J.~A.} \bibnamefont{Lotka}},
  \bibinfo{journal}{Journal of the American Chemical Society}
  \textbf{\bibinfo{volume}{42}}, \bibinfo{pages}{1595} (\bibinfo{year}{1920}).

\bibitem[{\citenamefont{Volterra}(1926)}]{Volterra1926}
\bibinfo{author}{\bibfnamefont{V.}~\bibnamefont{Volterra}},
  \bibinfo{journal}{Nature} \textbf{\bibinfo{volume}{118}},
  \bibinfo{pages}{558} (\bibinfo{year}{1926}).

\bibitem[{\citenamefont{Rasmussen et~al.}(2008)\citenamefont{Rasmussen, Bedau,
  Chen, Deamer, Krakauer, Packard, and Stadler}}]{Rasmussen2008}
\bibinfo{editor}{\bibfnamefont{S.}~\bibnamefont{Rasmussen}},
  \bibinfo{editor}{\bibfnamefont{M.~A.} \bibnamefont{Bedau}},
  \bibinfo{editor}{\bibfnamefont{L.}~\bibnamefont{Chen}},
  \bibinfo{editor}{\bibfnamefont{D.}~\bibnamefont{Deamer}},
  \bibinfo{editor}{\bibfnamefont{D.~C.} \bibnamefont{Krakauer}},
  \bibinfo{editor}{\bibfnamefont{N.~H.} \bibnamefont{Packard}},
  \bibnamefont{and} \bibinfo{editor}{\bibfnamefont{P.~F.}
  \bibnamefont{Stadler}}, eds., \emph{\bibinfo{title}{Protocells, Bridging
  Nonliving and Living Matter}} (\bibinfo{publisher}{MIT Press (Cambridge
  Mass.)}, \bibinfo{year}{2008}).

\bibitem[{\citenamefont{Luisi et~al.}(2006)\citenamefont{Luisi, Ferri, and
  Stano}}]{Luisi2006}
\bibinfo{author}{\bibfnamefont{P.~L.} \bibnamefont{Luisi}},
  \bibinfo{author}{\bibfnamefont{F.}~\bibnamefont{Ferri}}, \bibnamefont{and}
  \bibinfo{author}{\bibfnamefont{P.}~\bibnamefont{Stano}},
  \bibinfo{journal}{Naturwissenschaften} \textbf{\bibinfo{volume}{93}},
  \bibinfo{pages}{1} (\bibinfo{year}{2006}).

\bibitem[{\citenamefont{Szathmary}(2005)}]{Szath2005}
\bibinfo{author}{\bibfnamefont{E.}~\bibnamefont{Szathmary}},
  \bibinfo{journal}{Nature} \textbf{\bibinfo{volume}{433}},
  \bibinfo{pages}{469} (\bibinfo{year}{2005}).

\bibitem[{\citenamefont{Beck et~al.}(2011)\citenamefont{Beck, Topf, Frazier,
  Tjong, Xu, Zhang, and Alber}}]{Beck2011}
\bibinfo{author}{\bibfnamefont{M.}~\bibnamefont{Beck}},
  \bibinfo{author}{\bibfnamefont{M.}~\bibnamefont{Topf}},
  \bibinfo{author}{\bibfnamefont{Z.}~\bibnamefont{Frazier}},
  \bibinfo{author}{\bibfnamefont{H.}~\bibnamefont{Tjong}},
  \bibinfo{author}{\bibfnamefont{M.}~\bibnamefont{Xu}},
  \bibinfo{author}{\bibfnamefont{S.}~\bibnamefont{Zhang}}, \bibnamefont{and}
  \bibinfo{author}{\bibfnamefont{F.}~\bibnamefont{Alber}},
  \bibinfo{journal}{Journal of Structural Biology}
  \textbf{\bibinfo{volume}{173}}, \bibinfo{pages}{483} (\bibinfo{year}{2011}).

\bibitem[{\citenamefont{Gillespie et~al.}(2013)\citenamefont{Gillespie,
  Hellander, and Petzold}}]{Linda2013}
\bibinfo{author}{\bibfnamefont{D.~T.} \bibnamefont{Gillespie}},
  \bibinfo{author}{\bibfnamefont{A.}~\bibnamefont{Hellander}},
  \bibnamefont{and} \bibinfo{author}{\bibfnamefont{L.}~\bibnamefont{Petzold}},
  \bibinfo{journal}{The Journal of Chemical Physics}
  \textbf{\bibinfo{volume}{138}}, \bibinfo{pages}{170901}
  (\bibinfo{year}{2013}).

\bibitem[{\citenamefont{Andersen}(2005)}]{Andersen2005}
\bibinfo{author}{\bibfnamefont{O.~S.} \bibnamefont{Andersen}},
  \bibinfo{journal}{J Gen Physiol} \textbf{\bibinfo{volume}{125}},
  \bibinfo{pages}{3} (\bibinfo{year}{2005}).

\bibitem[{\citenamefont{Van~Kampen}(2007)}]{VanKampen2007}
\bibinfo{author}{\bibfnamefont{N.~G.} \bibnamefont{Van~Kampen}},
  \emph{\bibinfo{title}{Stochastic Processes in Physics and Chemistry}}
  (\bibinfo{publisher}{Elsevier}, \bibinfo{year}{2007}), \bibinfo{edition}{3rd}
  ed.

\bibitem[{\citenamefont{Schreiber}(2000)}]{Schreiber2000}
\bibinfo{author}{\bibfnamefont{T.}~\bibnamefont{Schreiber}},
  \bibinfo{journal}{Physical Review Letters} \textbf{\bibinfo{volume}{85}},
  \bibinfo{pages}{461} (\bibinfo{year}{2000}).

\bibitem[{\citenamefont{Hlavackova-Schindler
  et~al.}(2007)\citenamefont{Hlavackova-Schindler, Palus, Vejmelka, and
  Bhattacharya}}]{Palus2007}
\bibinfo{author}{\bibfnamefont{K.}~\bibnamefont{Hlavackova-Schindler}},
  \bibinfo{author}{\bibfnamefont{M.}~\bibnamefont{Palus}},
  \bibinfo{author}{\bibfnamefont{M.}~\bibnamefont{Vejmelka}}, \bibnamefont{and}
  \bibinfo{author}{\bibfnamefont{J.}~\bibnamefont{Bhattacharya}},
  \bibinfo{journal}{Physics Reports} \textbf{\bibinfo{volume}{441}},
  \bibinfo{pages}{1} (\bibinfo{year}{2007}).

\bibitem[{\citenamefont{Vejmelka and Palus}(2008)}]{Palus2008}
\bibinfo{author}{\bibfnamefont{M.}~\bibnamefont{Vejmelka}} \bibnamefont{and}
  \bibinfo{author}{\bibfnamefont{M.}~\bibnamefont{Palus}},
  \bibinfo{journal}{Phys Rev E Stat Nonlin Soft Matter Phys}
  \textbf{\bibinfo{volume}{77}}, \bibinfo{pages}{026214}
  (\bibinfo{year}{2008}).

\bibitem[{\citenamefont{Cover and Thomas}(2006)}]{Cover2006}
\bibinfo{author}{\bibfnamefont{T.~M.} \bibnamefont{Cover}} \bibnamefont{and}
  \bibinfo{author}{\bibfnamefont{J.~A.} \bibnamefont{Thomas}},
  \emph{\bibinfo{title}{Elements of information theory}}
  (\bibinfo{publisher}{Wiley-Interscience}, \bibinfo{address}{Hoboken, N.J.},
  \bibinfo{year}{2006}), \bibinfo{edition}{2nd} ed.

\bibitem[{\citenamefont{Powell and Percival}(1979)}]{Percival1979}
\bibinfo{author}{\bibfnamefont{G.~E.} \bibnamefont{Powell}} \bibnamefont{and}
  \bibinfo{author}{\bibfnamefont{I.~C.} \bibnamefont{Percival}},
  \bibinfo{journal}{Journal of Physics A: Math. Gen.}
  \textbf{\bibinfo{volume}{12}}, \bibinfo{pages}{2053} (\bibinfo{year}{1979}).

\bibitem[{\citenamefont{Rosso et~al.}(2007)\citenamefont{Rosso, Larrondo, T,
  Plastino, and Fuentes}}]{Fuentes2007}
\bibinfo{author}{\bibfnamefont{O.~A.} \bibnamefont{Rosso}},
  \bibinfo{author}{\bibfnamefont{H.~A.} \bibnamefont{Larrondo}},
  \bibinfo{author}{\bibfnamefont{M.~M.} \bibnamefont{T}},
  \bibinfo{author}{\bibfnamefont{A.}~\bibnamefont{Plastino}}, \bibnamefont{and}
  \bibinfo{author}{\bibfnamefont{M.~A.} \bibnamefont{Fuentes}},
  \bibinfo{journal}{Physical Review Letters} \textbf{\bibinfo{volume}{99}},
  \bibinfo{pages}{154102} (\bibinfo{year}{2007}).

\bibitem[{\citenamefont{Ester et~al.}(1996)\citenamefont{Ester, Kriegel,
  Sander, and Xiaowei}}]{Ester1996}
\bibinfo{author}{\bibfnamefont{M.}~\bibnamefont{Ester}},
  \bibinfo{author}{\bibfnamefont{H.-P.} \bibnamefont{Kriegel}},
  \bibinfo{author}{\bibfnamefont{J.}~\bibnamefont{Sander}}, \bibnamefont{and}
  \bibinfo{author}{\bibfnamefont{X.}~\bibnamefont{Xiaowei}}, in
  \emph{\bibinfo{booktitle}{Proceedings of the Second International Conference
  on Knowledge Discovery and Data Mining (KDD-96)}}, edited by
  \bibinfo{editor}{\bibfnamefont{E.}~\bibnamefont{Simoudis}},
  \bibinfo{editor}{\bibfnamefont{J.}~\bibnamefont{Han}}, \bibnamefont{and}
  \bibinfo{editor}{\bibfnamefont{U.~M.} \bibnamefont{Fayyad}}
  (\bibinfo{year}{1996}).

\end{thebibliography}

\newpage

\begin{table}[!h]
\begin{center}
\caption{Mutual information between population size and largest cluster size for species $A$, $B$ and $C$ and for different numbers of inert crowders.} 
\label{tab:1}
\begin{tabular}{|c||c|c|c|}
      \hline
      Inert Crowders & A & B & C\\
      \hline
      0 & 0.969 & 0.811 & 1.574 \\
      \hline
      2000 & 0.893 & 0.643 & 1.359 \\
      \hline
      4000 & 0.847 & 0.619 & 0.806 \\
      \hline
      6000 & 0.611 & 0.583 & 0.735 \\
      \hline
      8000 & 0.568 & 0.471 & 0.442 \\
      \hline
\end{tabular}
\end{center}
\end{table}

\newpage

\newpage

\newpage

\newpage

\newpage

\newpage

\newpage

\end{document}